\documentclass{PoS}

\title{CASA on the fringe: VLBI data processing in the CASA software package}

\ShortTitle{CASA on the fringe}

\author{\speaker{Ilse van Bemmel}, Des Small, Mark Kettenis, Arpad Szomoru\\
    Joint Institute for VLBI ERIC (JIVE)\\
    Oude Hoogeveensedijk 4, 7991\,PD Dwingeloo, The Netherlands\\
    E-mail: \email{bemmel@jive.eu}}

\author{George Moellenbrock\\
    National Radio Astronomy Observatory (NRAO)\\
    NRAO Array Operations Center, P.O.Box O, Socorro, NM, USA}    

\author{Michael Janssen\\
    IMAPP - Radboud University\\
    Faculty of Science, P.O. Box 9010, 6500\,GL Nijmegen, The Netherlands}

\abstract{In recent years new functionality for VLBI data processing has been added to the CASA package. This paper presents the new CASA tasks {\em fringefit} and {\em accor}, which are closely matched to their AIPS counterparts FRING and ACCOR. Several CASA tasks received upgrades to handle VLBI specific metadata. With the current CASA release VLBI data processing is possible, and functionality will be expanded in the upcoming release. Longer term developments include fringe fitting of broad, non-continuous frequency bands and dispersive delays, which will ensure that the number of use cases for VLBI calibration will increase in future CASA releases.}

\FullConference{14th European VLBI Network Symposium \& Users Meeting (EVN 2018)\\
		8-11 October 2018\\
		Granada, Spain}

\begin{document}

\section{Introduction}
In recent years, the CASA software package \cite{CASA} has become the package of choice to process and calibrate the majority of radio data, and all early career radio astronomers are now trained to use this package. However, for Very Long Baseline Interferometry (VLBI) data observers have continued to use AIPS \cite{aips}. Until recently CASA lacked functionality to handle VLBI data properly, such as a fringe fitting task, and adequate support for handling the VLBI (meta-)data. With support from the BlackHoleCam \cite{BHC} and SKA-NL projects, JIVE has developed the new CASA tasks {\em fringefit} and {\em accor}, and upgraded other tasks to work properly with VLBI observations. In this paper we describe the development and first experimental CASA release of the new and upgraded VLBI functionality.

\section{Fringe fitting basics}
Fringe fitting is a crucial step in VLBI data processing, that is not commonly used for data taken with connected element arrays, which share a clock and view of the atmosphere. After correlation the VLBI data still contain small phase errors due to instrumental and atmospheric effects. These appear as phase offsets between spectral windows, or slopes in phase as a function of time or frequency. Due to the long baselines in VLBI, these slopes can be very steep, and cause coherence loss when averaging the data in time and/or frequency. 

The fringe fitting process corrects the visibility phases for these problems, both in time and frequency. It solves for phase offsets, delay (phase slope as function of frequency) and rate (phase slope as function of time). The solutions are per antenna, ensuring that phase closure information is not affected. For the majority of VLBI observations, fringe fitting is a two-step process, first removing the instrumental effects, and then solving for time variable atmospheric issues. Depending on instrument and frequency, additional steps or different order of the steps may be needed.
Instrumental effects are usually stable over an observation, and a single bright calibrator scan can be used to determine the instrumental delay and phase offsets between spectral windows{\footnote{A spectral window in CASA is the equivalent of an AIPS IF}}.
For atmospheric effects a higher time cadence is needed and also solutions for the rate need to be found. This requires the source to be bright enough to be detected in a short time and narrow frequency band, which is usually not the case for the target source. To circumvent this, the technique of phase referencing uses a bright calibrator close to the target source. The target and calibrator are observed in an alternating sequence of scans, and fringe fitting is done on the calibrator scans. The cadence is chosen to match the atmospheric coherence time at the frequency of the observations, typically of order 10 minutes for 1.5\,GHz. This allows for interpolation of the fringe fitting solutions between calibrator scans, and apply them to the target source. For cm-wavelength VLBI this well established technique enables deep observations of faint sources. 

For current and planned VLBI facilities, the AIPS packages becomes ever more challenging to maintain. The FORTRAN base has limited flexibility to cover all possible use cases. Interoperability with packages using Measurement Sets is also challenging. To secure processing of VLBI data in the future, and expand the usability to new applications such as Event Horizon Telescope, or even the Square Kilometre Array, it was found that CASA would offer the best opportunities. The approach to making CASA ready for VLBI was to stay as close as possible to the existing functionality in AIPS. This enables direct comparison of the tasks with simulated and real data, and makes the look and feel of tasks, as well as the work-flow, familiar to AIPS users.

\section{Development}
At JIVE a pipeline is used for data quality checks of observations from the European VLBI Network (EVN). It runs with AIPS ParselTongue \cite{parseltongue} and provides all basic steps required for VLBI data processing. As such it forms a minimal set of tasks we need to have present in CASA. After identifying the comparable tasks in CASA it was evident that new metadata handling and fringe fitting were the two main hurdles for processing VLBI data. 

The new {\em fringefit} task implements the Schwab-Cotton global fringe fitting algorithm \cite{schwabcotton}, and is comparable to the AIPS task FRING. Development started with building a Python prototype, which was extensively tested against FRING. The results from the prototype and FRING were indistinguishable. As expected, in the chosen format the prototype performed slower than AIPS FRING.

The production code is fully integrated in the CASA framework, such that all methods for data selection and a-priori calibration are available to the user. The C++ implementation is also significantly faster than the prototype, though AIPS FRING is still a factor of 2-4 faster. The parallelization options in CASA make it easy to overcome this difference in processing speed.

\begin{figure}[!t]
    \centering
    \includegraphics[width=7.5cm]{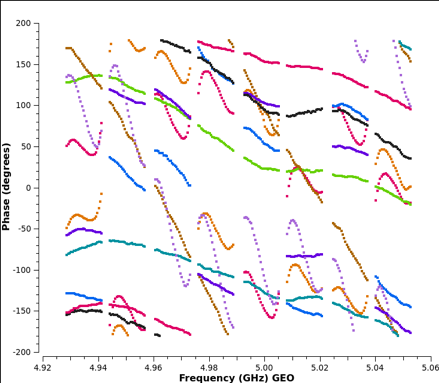}
    \includegraphics[width=7.5cm]{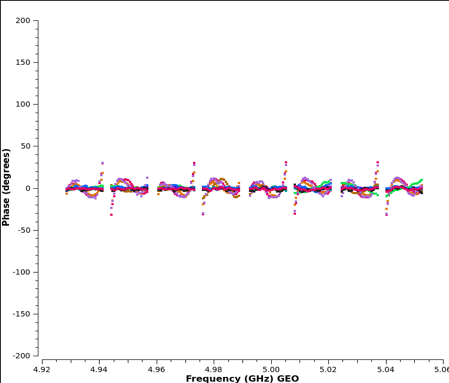}
    \caption{Phases as a function of frequency demonstrate the impact of instrumental phase and delay corrections on an observation of a bright calibrator source. The colour coding is for different baselines, all to the reference antenna. Left is before, right is after correction. The residual phase effects in the right plot are due to non-linear phase bandpass which is not corrected.}
    \label{SBD}
\end{figure}

Figure~\ref{SBD} shows the results of instrumental phase and delay corrections on an EVN scan of a bright calibrator source. Non-linear bandpass phase residuals remain evident, and may be calibrated using the standard CASA bandpass task.

With different correlators used in VLBI arrays, some additional tasks were needed to ensure that all data is properly handled in CASA. The new task {\em accor} is the equivalent of AIPS ACCOR, and has been developed to normalize the visibilities by the auto-correlation amplitudes. To import the FITS-IDI data with metadata included, improvements were made to the {\em importfitsidi} task. Changes were also made to apply digital corrections for the DiFX correlator.

The native data format in CASA is Measurement Set (MS), while the EVN archive supplies FITS-IDI data with additional metadata stored in UVFLG and ANTAB files. This implies that some translation scripts are needed to ensure that CASA can handle the metadata. The system temperature information from the ANTAB file has to be appended to the FITS-IDI file. The gain-curve information from the ANTAB file also needs to be interpreted and reformatted for importing into CASA. The UVFLG tables have to be translated into a CASA readable flag file. These steps are supported with Python scripts. For each dataset obtained from the EVN archive these steps only need to be run once.

\section{Public releases}
After extensive in-house testing, the software was at a level to be tested by experts in October 2017. During a RadioNet sponsored workshop $\sim20$ people were invited to test the new VLBI functionality on observations from a range of VLBI arrays, including EVN, VLBA, LBA, KVN, LOFAR, e-MERLIN, EHT, and even RadioAstron. The workshop uncovered many issues that were fixed on the fly, or in later development. The code has since matured further, and is included as an experimental release in CASA 5.3 and 5.4. Only basic, essential functionality is included; the task does not yet match the full functionality of AIPS FRING. Development is ongoing, and several improvements are planned for CASA 5.5 release in the spring of 2019. 

With the release of CASA 5.3 JIVE is actively engaging the VLBI community in the further development of the VLBI functionality. EVN data has been processed and imaged successfully, and e-MERLIN, LOFAR and EHT are exploring the new CASA tasks for their data processing. 

\begin{figure}[!t]
    \centering
    \includegraphics[width=7.5cm]{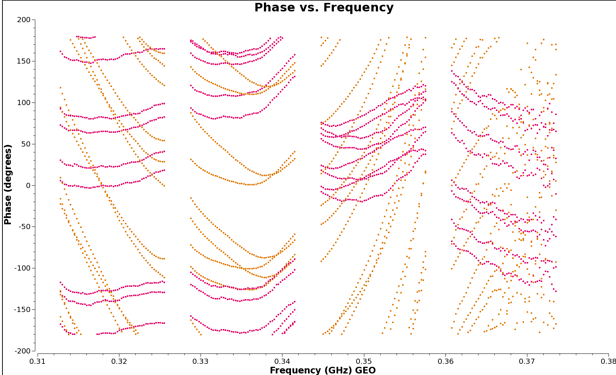}
    \includegraphics[width=7.5cm]{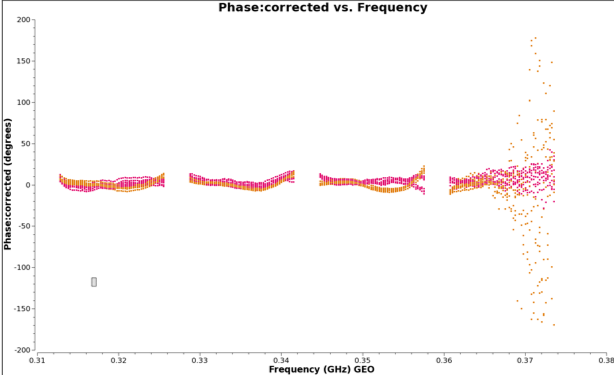}
    \caption{Phase as a function of frequency demonstrates the effect of dispersive delay correction on EVN P-band data (350\,GHz). Left is before and right is after correction. Multiple baselines to the reference antenna are shown, colour coding is for RCP and LCP. The dispersive delay is clearly visible as an additional curvature in the phases before calibration, which is removed by the fringe fitting. At the high frequency end the data quality deteriorates, causing a clear increase in noise.}
    \label{dispersivedelay}
\end{figure}

\section{Ongoing and future work}
Development of the CASA VLBI functionality is continuing under the RadioNet RINGS project, and further improvements are expected to appear in future CASA releases. The longer term development will include fringe fitting of broad and/or non-contiguous frequency bands, as well as correction for dispersive delays. A prototype for the dispersive delay calibration is operational, as shown in Figure~\ref{dispersivedelay}.
Efforts to develop pipelines for VLBI data-processing in CASA are also taking off. The rPICARD pipeline \cite{rpicard} is currently used for data processing of Event Horizon Telescope observations, and its products are part of the first science release. E-MERLIN uses CASA for its pipeline, and EVN pipeline development is starting. The planned improvements to the CASA VLBI software and move of CASA to Python 3 (CASA 6, expected mid-2019), will allow for many more use cases to exploit CASA.
As the VLBI functionality is still actively developed, we solicit feedback from users through the NRAO Helpdesk system.

\subsection*{Acknowledgements}
The authors would like to thank Anita Richards for being one of the earliest adopters of CASA for VLBI data processing. Also thanks to the participants of the CASA workshop at JIVE, and a special mention to Ivan Marti-Vidal for discussions on calibration in general and fringe fitting in particular. Lastly, thanks to Dirk Petry for his work on the {\em importfitsidi} task.
This work is supported by the ERC Synergy Grant "BlackHoleCam: Imaging the Event Horizon of Black Holes" (Grant 610058). This work was also supported by SKA-NL. The workshop hosted at JIVE has received funding from the European Union's Horizon 2020 research and innovation programme RadioNet under grant agreement No 730562. The RINGS project is a RadioNet Joint Research Activity. The National Radio Astronomy Observatory is a facility of the National Science Foundation operated under cooperative agreement by Associated Universities, Inc.

\end{document}